\documentclass[jkps,preprint,fleqn,showpacs,showkeys,10pt]{revtex4}
\usepackage{graphicx}
\usepackage{amssymb}
\usepackage{amsmath}
\usepackage{bm}

\usepackage{latexsym,amsfonts}
\usepackage[dvipsnames]{xcolor}
\usepackage{epsfig}
\usepackage[all]{xy}
\begin{document}

\setcounter{page}{0}
\title[]{Demonstration of the Hayden-Preskill Protocol via Mutual Information}
\author{Jeong-Myeong \surname{Bae}}
\email{bjmhk2@dgist.ac.kr}
\affiliation{School of Undergraduate Studies, College of Transdisciplinary Studies, 
Daegu Gyeongbuk Institute of Science and Technology (DGIST), Daegu 42988, Republic of Korea}
\author{Subeom \surname{Kang}}
\email{ksb527@kaist.ac.kr}
\affiliation{Department of Physics, KAIST, Daejeon 34141, Republic of Korea}
\author{Dong-han \surname{Yeom}}
\email{innocent.yeom@gmail.com}
\affiliation{Department of Physics Education, Pusan National University, Busan 46241, Republic of Korea}
\affiliation{Research Center for Dielectric and Advanced Matter Physics, Pusan National University, Busan 46241, Republic of Korea}
\author{Heeseung \surname{Zoe}}
\email{heezoe@dgist.ac.kr}
\affiliation{School of Undergraduate Studies, College of Transdisciplinary Studies,  
Daegu Gyeongbuk Institute of Science and Technology (DGIST), Daegu 42988, Republic of Korea}


\begin{abstract}
We construct the Hayden-Preskill protocol by using a system of spin-1/2 particles and demonstrate information flows of this system which can mimic black holes. We first define an analogous black hole $A$ as a collection of such particles. Second, we take the particles from inside to outside the black hole to define an analogous system of Hawking radiation $B$ as outside particles. When the black hole and the radiation have the maximum entanglement at the Page time, we take an entangled pair system $C$ and $D$. The particles of $C$ fall into the black hole while their counterparts of $D$ remain outside. If we assume rapid mixing of the particle states in the black hole $A \cup C$, can the information of $C$ rapidly escape from the black hole like a mirror? We numerically show that if we turn on the rapid mixing in the black hole, the original information of $C$ rapidly escapes from the black hole to outside in the form of the mutual information between $B$ and $D$. On the other hand, if the mixing between $A$ and $C$ is not enough, the information escapes slowly. Hence, we explicitly demonstrate the original conjecture of Hayden and Preskill. We emphasize that enough mixing is an essential condition to make the Hayden-Preskill protocol functionally work.
\end{abstract}



\maketitle

\section{Introduction}

Constructing toy models reflecting essential features of quantum mechanics can be useful for tackling the information loss problem of black holes \cite{Hawking:1976ra}. If we present a black hole system as a collection of spin-1/2 particles, we can describe the information flows between the parts of the system straightforwardly \cite{Page:1993df,Hwang:2016otg,Hwang:2017yxp}. Generally, we define information as the difference between the Boltzmann entropy and the entanglement entropy, and it measures the bias from the exact thermal state \cite{Lloyd:1988cn}. 
As a simple example, we consider a black hole as a bipartite system consisting of a black hole and radiation. In this case, one can fully understand the information flow by calculating the information of the black hole, the information of the radiation, and the mutual information between the black hole and the radiation. The total sum of information is preserved \cite{Alonso-Serrano:2015bcr}. If we further assume that the states are pure and random, the information of radiation is then negligible by the time the initial Boltzmann entropy of the black hole has decreased to half its value. We call this moment the Page time \cite{Page:1993df,Hwang:2016otg}.

If the Boltzmann entropy is proportional to the areal radius, then the information should be attached by Hawking radiation, because the black hole will still be semi-classical around the Page time \cite{Susskind:1993if}. However, if a Hawking particle carries information, then it causes more severe problems \cite{Yeom:2008qw,Yeom:2009zp,Almheiri:2012rt}. That some researchers consider various kinds of remnants that carry all information is not surprising \cite{Chen:2014jwq}. The exact role of the remnant in terms of entanglements and information can be well investigated by using the spin-1/2 toy model \cite{Hwang:2016otg}, where now we need to introduce three parts: the black hole, the radiation, and the remnant. From this toy model, we can obtain several interesting results; for example, the information locking scenario cannot be realized unless the Boltzmann entropy of the final stage of the black hole is very huge \cite{Smolin:2005tz}.

The tension between semi-classical gravity and the unitarity of quantum mechanics can be demonstrated by using a spin-1/2 toy model \cite{Hwang:2017yxp}. The idea that Hawking radiation is generated at the pair creation of two particles around the horizon is reasonable. Note that these two particles are separable from the black hole and the radiation that exist before the pair creation. The number of total states must increase linearly as the pair particles are added. Hence, an annihilation process should take place inside the black hole to prevent this situation. To make this process unitarily, the infalling antiparticle should find a partner particle to be annihilated inside the black hole. This new antiparticle-particle pair should form a new separable state. Interestingly, from the calculation of the toy model, one can show that such a unitary annihilation process is impossible after the Page time \cite{Hwang:2017yxp}. One possible interpretation is that the Hawking particle emission process is not equivalent to the creation of separable pair particles, where this violates the no drama condition near the horizon \cite{Almheiri:2012rt} which can be potentially very harmful \cite{Hwang:2012nn}. Another interpretation is that the number of states inside a black hole should linearly increase, at least after the Page time \cite{Hwang:2017yxp}, although this conclusion is also unsatisfactory \cite{Giddings:1994qt}.

In the sense that the lessons from the spin-1/2 particles show the difficulty of the information loss paradox, they are fruitful and educative. In this context, we will demonstrate one other important idea of the information theoretical issue of the black hole evaporation, which is known by the \textit{Hayden-Preskill protocol} \cite{Hayden:2007cs}. According to Hayden and Preskill, a small bit of information, which is inserted to the black hole after the Page time, will be rapidly emitted from the black hole like a mirror reflection. Here, the rapidness is related to the time scale of sufficient randomization or scrambling. The time scale of scrambling is believed to be marginally consistent with the black hole complementarity principle. However, later, one can show that such a marginal bound for black hole complementarity can breakdown \cite{Yeom:2009zp}. Moreover, if the Hayden-Preskill protocol is true, one can show that the Horowitz-Maldacena proposal \cite{Horowitz:2003he}, which is a candidate resolution of the information loss problem, may still be an unsatisfactory idea \cite{Hong:2008ga}. On the other hand, if the Hayden-Preskill protocol is not true, some of the interesting previous studies will lose their own grounds.

Therefore, the explicit demonstration of the Hayden-Preskill protocol is an important and interesting issue that we need to investigate. In Section~\ref{sec:dem}, we first illustrate the original version of the Hayden-Preskill protocol and second show the model with spin-1/2 particles. In Section~\ref{sec:num}, we show several numerical experiments and confirm that the original Hayden-Preskill conjecture functionally works as long as we introduce a rapid mixing. Finally, in Section~\ref{sec:con}, we critically review this paper and discuss possible future applications.

\section{\label{sec:dem}Demonstrating the Hayden-Preskill Protocol}

In this section, we first summarize the original argument and theoretical justification of Hayden and Preskill \cite{Hayden:2007cs}. After this, we illustrate a way to demonstrate the Hayden-Preskill argument by using spin-1/2 particles.

\subsection{Review of the Hayden-Preskill Protocol}

Hayden and Preskill \cite{Hayden:2007cs} considered a situation oh which a black hole first exists after the Page time. The total system is the black hole (say, $A$) and radiation (say, $B$), where the number of states for $A$ is smaller than that for $B$. Next, one puts a small bit of information, say $C$, into the black hole. Finally, one tests when the infalling information will escape from the black hole.

The next natural question is how to confirm that the emitted particle carries information about $C$. Hayden and Preskill introduce a particle $D$, which forms a maximally entangled pair with $C$. While $C$ falls into the black hole, $D$ remains outside. Because $C$ and $D$ are maximally entangled, one can check whether the emitted information is about $C$ or not, by comparing the state with $D$ and the radiation. More rigorously, one can prove that the fidelity about $D$ is almost perfectly secured if the system is assumed to be random. Hayden and Preskill named this behavior \textit{black holes as mirrors}.

A key feature of the Hayden-Preskill protocol is the randomizing process applied to the black hole $A$ and the infalling matter $C$ after the Page time. However, time is needed to randomize $A$ and  $C$, and time might have a fundamental limitation. Based on simple arguments, they proposed that the minimum time scale is about $\sim T^{-1} \log S$, where $T$ is the temperature and $S$ is the entropy. In terms of the black hole mass, it becomes $\sim M \log M$, which they call the scrambling time.

When the paper of Hayden and Preskill was published, they considered this behavior in the context of black hole complementarity \cite{Susskind:1993if}, which says that at least two complementary observers exist, where one is an asymptotic observer and the other is an infalling observer. According to the asymptotic observer, the black hole is a kind of membrane, and all infalling information attached to the horizon is thermalized and emitted via Hawking radiation. Because all the processes are causally connected, the asymptotic observer sees no information loss. However, if the information is emitted too rapidly, the asymptotic observer can compare the information with the infalling observer, which can cause a severe inconsistency \cite{Susskind:1993mu}. Hayden and Preskill reported that the minimum time scale is $\sim M \log M$, which is marginally consistent with the black hole complementarity principle.

However, by increasing the number of matter fields that contribute to Hawking radiation, one can explicitly show that this marginal bound can be irrelevant within the semi-classical regime. This criticism is consistent with \cite{Almheiri:2012rt}. It strongly implies that now the original motivation of the Hayden-Preskill protocol is lost. If no membrane exists at the event horizon, then Hawking radiation from the black hole is not equivalent to particles taken directly from $A \cup C$  \cite{Hwang:2017yxp}. Rather, we should introduce a separable particle-antiparticle pair generated around the horizon. While the outgoing particle becomes Hawking radiation, the antiparticle falls into the black hole.  Then can the Hayden-Preskill protocol still work even without using black hole complementarity or the membrane paradigm \cite{Thorne:1986iy}? This goes beyond the scope of this paper, but we will come back to this question later.

\subsection{Modeling by Spin-1/2 Particles}

In order to demonstrate the Hayden-Preskill protocol using spin-1/2 particles, we define the following four subsystems:
\begin{itemize}
\item[--] 1. Black hole: $A$, where the number of particles is $N_{A}$ and the number of states is $a = 2^{N_{A}}$.
\item[--] 2. Radiation: $B$, where the number of particles is $N_{B}$ and the number of states is $b = 2^{N_{B}}$.
\item[--] 3. Infalling information: $C$, where the number of particles is $N_{C}$ and the number of states is $c = 2^{N_{C}}$.
\item[--] 4. Reference of $C$: $D$, where the number of particles is $N_{D}$ and the number of states is $d = 2^{N_{D}}$.
\end{itemize}

\subsubsection{Black hole and radiation: a bipartite system}

As an initial condition, we assume that $A$ and $B$ are randomly mixed and that $N_{A} + N_{B} = \mathrm{const.}$ In the beginning, $N_{B}=0$ and as time goes on $N_{B}$ linearly increases. The quantum state is
\begin{eqnarray}
| \psi_{A\cup B} \rangle = \sum_{i_{1}, ... ,i_{N_{A}+N_{B}}} c_{i_{1}, ... , i_{N_{A}+N_{B}}} | i_{1}, ... , i_{N_{A}+N_{B}} \rangle,
\end{eqnarray}
where $i_{j} = 1, 2$ (up or down), $| i_{1}, ... , i_{N_{A}+N_{B}} \rangle$ are an orthonormal basis, and $c_{i_{1}, ... , i_{N_{A}+N_{B}}}$ are complex numbers that satisfy the orthonormalization condition. By introducing random numbers $c_{i_{1}, ... , i_{N_{A}+N_{B}}}$, we can prepare a random state. By using this state, we can define the density matrix
\begin{eqnarray}
\rho_{A \cup B} = | \psi_{A\cup B} \rangle \langle \psi_{A\cup B} |.
\end{eqnarray}
We can further trace-out the degrees of freedom for each part, e.g.,
\begin{eqnarray}
\rho_{A} &=& \mathrm{tr}_{B} \rho_{A \cup B},\\
\rho_{B} &=& \mathrm{tr}_{A} \rho_{A \cup B}.
\end{eqnarray}
By using this, we can define the entanglement entropy
\begin{eqnarray}
S(A|B) = - \mathrm{tr} \rho_{A} \log \rho_{A}
\end{eqnarray}
and $S(A|B) = S(B|A)$ for a pure state.

From these definitions, we can define information for each part as follows:
\begin{itemize}
\item[--] Information for the black hole: $I_{A} \equiv \log a - S(A|B)$.
\item[--] Information for the radiation: $I_{B} \equiv \log b - S(B|A)$.
\item[--] Mutual information between $A$ and $B$: $I_{AB} \equiv S(A|B) + S(B|A) - S(A\cup B) = 2 S(A|B)$, where the last equality is derived assuming a pure state.
\end{itemize}
$I_{A}$ and $I_{B}$ show bias from the exact thermal state; hence, if $I_{A}$ or $I_{B}$ is greater than zero, then in principle, non-trivial information from the black hole or the Hawking radiation can be measured. Also, $I_{A} + I_{B} + I_{AB} = \mathrm{const.}$ can be shown, if the total number of particles is preserved.

\subsubsection{Inserting entangled particles}

Now, we need to prepare an entangled state of $C$ and $D$  and put $C$ into the black hole. If $C$ and $D$ are maximally entangled, the mutual information between $C$ and $D$ will be the largest. However, practically there is no physical difference, even when we consider a randomly entangled $C$ and $D$. Following the same path as in the previous subsection, we can define
\begin{eqnarray}
| \psi_{C\cup D} \rangle = \sum_{i_{1}, ... ,i_{N_{C}+N_{D}}} c_{i_{1}, ... , i_{N_{C}+N_{D}}} | i_{1}, ... , i_{N_{C}+N_{D}} \rangle.
\end{eqnarray}
In order to insert $C$ into $A$, we first consider the total state
\begin{eqnarray}
| \psi \rangle = | \psi_{A \cup B} \rangle \otimes | \psi_{C \cup D} \rangle.
\end{eqnarray}
On first glimpse, $A \cup B$ and $C \cup D$ are not entangled. However, we can introduce a randomizing interaction between $A$ and $C$, for example, by using the operation $\mathcal{O}^{\alpha\beta}$:
\begin{eqnarray}
\mathcal{O}^{\alpha\beta} \equiv \mathcal{S}^{\alpha\beta}_{A\cup C} \otimes \mathcal{I}_{B \cup D}, 
\end{eqnarray}
where $\alpha$ and $\beta$ are randomly chosen indices of $A$ and $C$, respectively, and $\mathcal{S}^{\alpha\beta}_{A\cup C}$ is the usual swap gate between $A$ and $C$. Repeating many operations, we can randomize the black hole states $A \cup C$.

Finally, we define $A' \equiv A \cup C$. The density matrix has effectively three parts: $A'$, $B$, and $D$. By tracing-out degrees of freedom, for this tripartite system, we can again define the information contents as follows:
\begin{itemize}
\item[--] Information for the black hole: $I_{A'} \equiv \log ac - S(A'|B \cup D)$.
\item[--] Information for the radiation: $I_{B} \equiv \log b - S(B|A' \cup D)$.
\item[--] Information for the reference: $I_{D} \equiv \log d - S(D|A' \cup C)$.
\item[--] Mutual information between $A'$ and $B$: $I_{A'B} \equiv S(A'|B \cup D) + S(B|A' \cup D) - S(A' \cup B | D)$.
\item[--] Mutual information between $B$ and $D$: $I_{BD} \equiv S(B|D \cup A') + S(D|B \cup A') - S(B \cup D | A')$.
\item[--] Mutual information between $D$ and $A'$: $I_{DA'} \equiv S(D|A' \cup B) + S(A'|D \cup B) - S(D \cup A' | B)$.
\item[--] Tripartite information: $I_{A'BD} \equiv S(A'|B \cup D) + S(B|D \cup A') + S(D|A' \cup B) - S(A' \cup B | D) - S(B \cup D|A') - S(D \cup A' | B) + S(A' \cup B \cup D)$.
\end{itemize}
Because the total system is a pure state, the sum of all information is preserved and the tripartite information must vanish \cite{Hwang:2016otg}.

As we take particles from $A \cup C$ to $B$, all information will be reduced to $I_{B}$, $I_{D}$, or $I_{BD}$. Because $C$ was entangled with $D$, the information about $C$ can be measured from the mutual information with $D$, i.e., $I_{BD}$. Therefore, by measuring $I_{BD}$, we can check whether the information about $C$ is emitted or not.

\section{\label{sec:num}Numerical Results}

In this section, we summarize our numerical results.

\begin{figure}
\begin{center}
\includegraphics[scale=0.4]{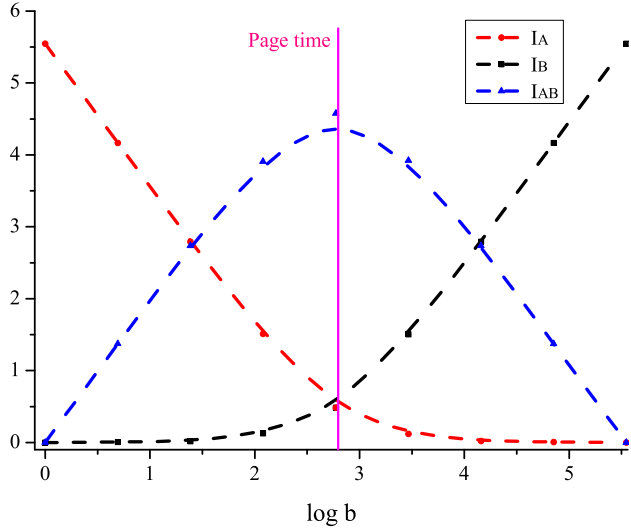}
\caption{\label{fig:data0}$I_{A}$, $I_{B}$, and $I_{AB}$.}
\end{center}
\end{figure}

\subsection{Null Result: Information Flow of a Bipartite System}

First, as a demonstration, we report the null result, where there is no $C$ and $D$. In Fig.~\ref{fig:data0}, the total number of particles is $8$. Note that, for all simulations in this paper, due to the random number dependence, we repeated 30 times and the results averaged. In addition, we connected dots smoothly by using B-splines to show a clear physical tendency.

As it has already been observed \cite{Page:1993df,Hwang:2016otg,Hwang:2017yxp}, the entanglement entropy follows the relation
\begin{eqnarray}
S(A|B) \simeq \log a
\end{eqnarray}
for $a < b$; if $b > a$, we can rely on the relation $S(A|B) = S(B|A)$. Before the Page time, the information of the black hole $I_{A}$ decreases and the mutual information between the black hole and radiation $I_{AB}$ increases as time goes on while the information of the radiation $I_{B}$ is negligible. After the Page time, $I_{B}$ increases while $I_{A}$ is negligible and $I_{AB}$ decreases. The sum of all information components is preserved, as we expected.

\begin{figure}
\begin{center}
\includegraphics[scale=0.4]{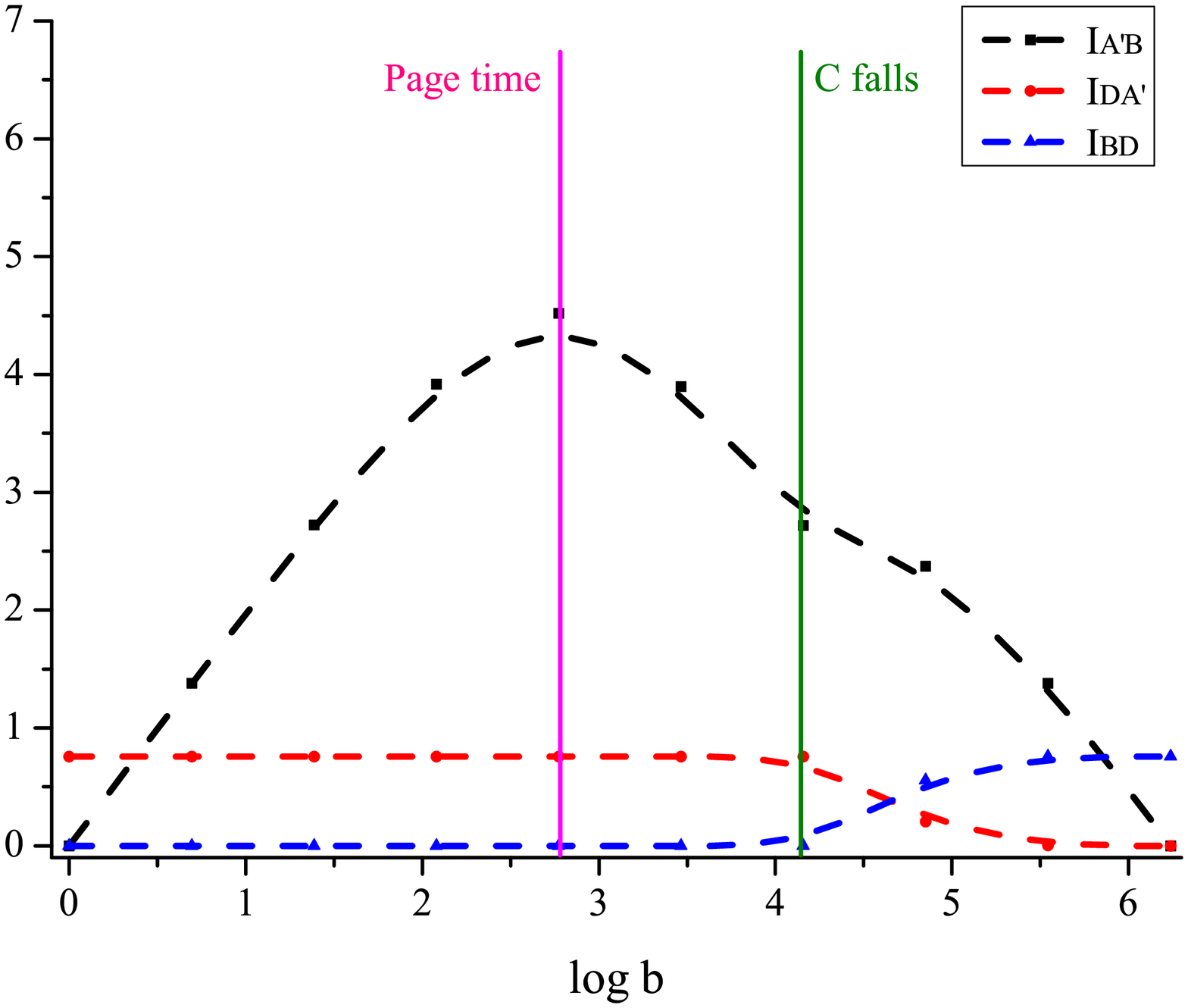}
\includegraphics[scale=0.4]{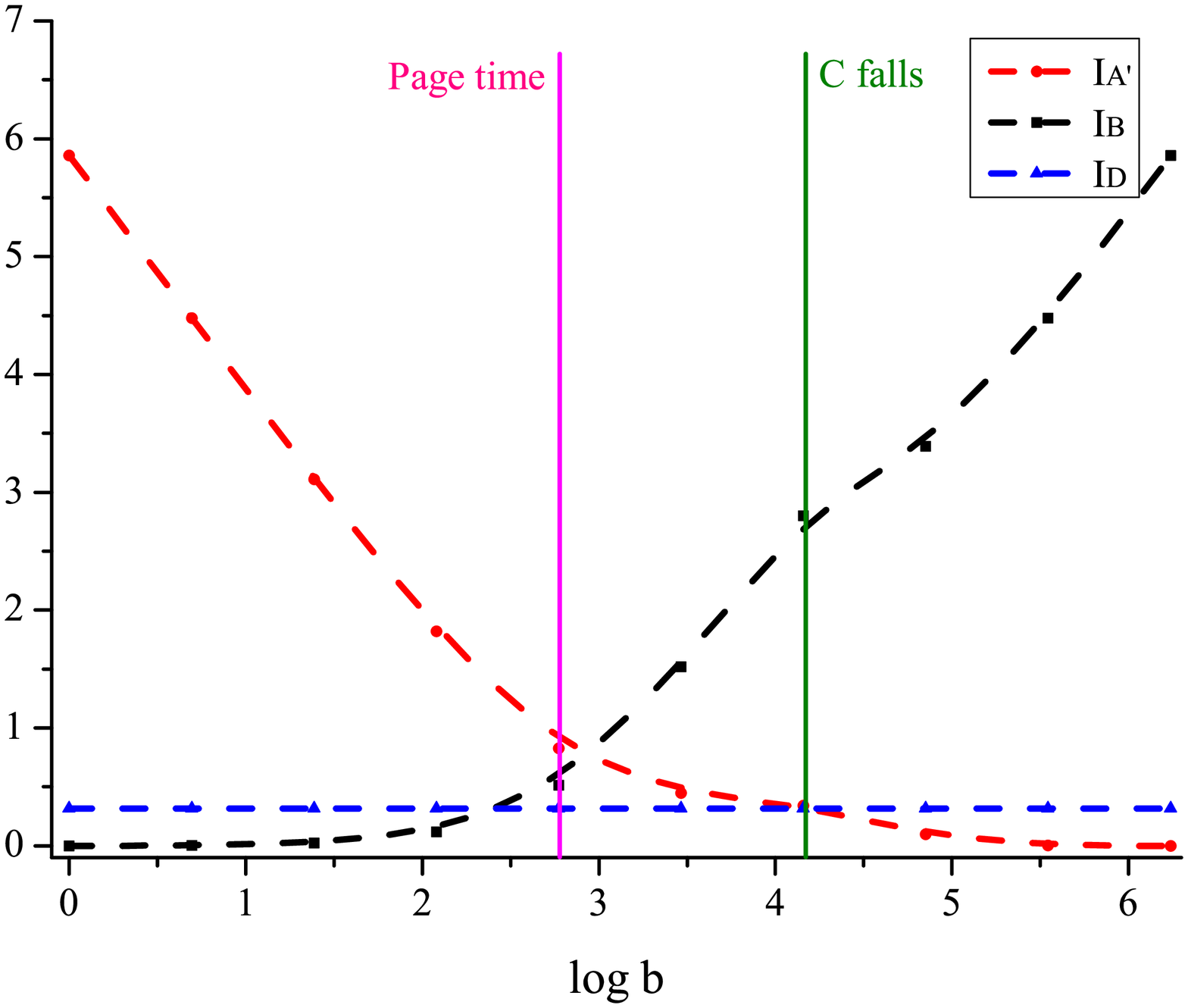}
\caption{\label{fig:data12}Numerical demonstration of the Hayden-Preskill protocol: upper: $I_{A'B}$, $I_{BD}$, and $I_{DA'}$ and lower: $I_{A'}$, $I_{B}$, and $I_{D}$.}
\end{center}
\end{figure}

\subsection{Rapid Information Emission after the Page Time}

Now, we insert $C$ into $A$ after the Page time. In Fig.~\ref{fig:data12}, we report the result, where $C$ and $D$ are the states of one particle. Initially, the black hole contains  8 particles. When 4 particles come out from the black hole, it reaches the Page time (pink lines in Fig.~\ref{fig:data12}).  After the Page time, we insert $C$ into the black hole while the entanglement entropy is decreasing  (say, 6 particles escaped)(green lines in Fig.~\ref{fig:data12}). Then, we turn on the mixing between $A$ and $C$ by using the swap gate. After 100 operations, we continue to take out particles to $B$.

Here, one technical note is that we add $C$ and mix it with $A$, where the original particle $C$ will be removed only in the last stage. If it is the black hole system, then the number of particles of $A$ is much larger than that of $C$; hence, probabilistically, the particles of $C$ will not contribute thoughtfully. However, in our toy model, the probability of choosing $C$ is quite high, which will significantly affect the final result and will spoil our interest in checking whether the mutual information rapidly carries the original information or not.

\begin{figure}
\begin{center}
\includegraphics[scale=0.4]{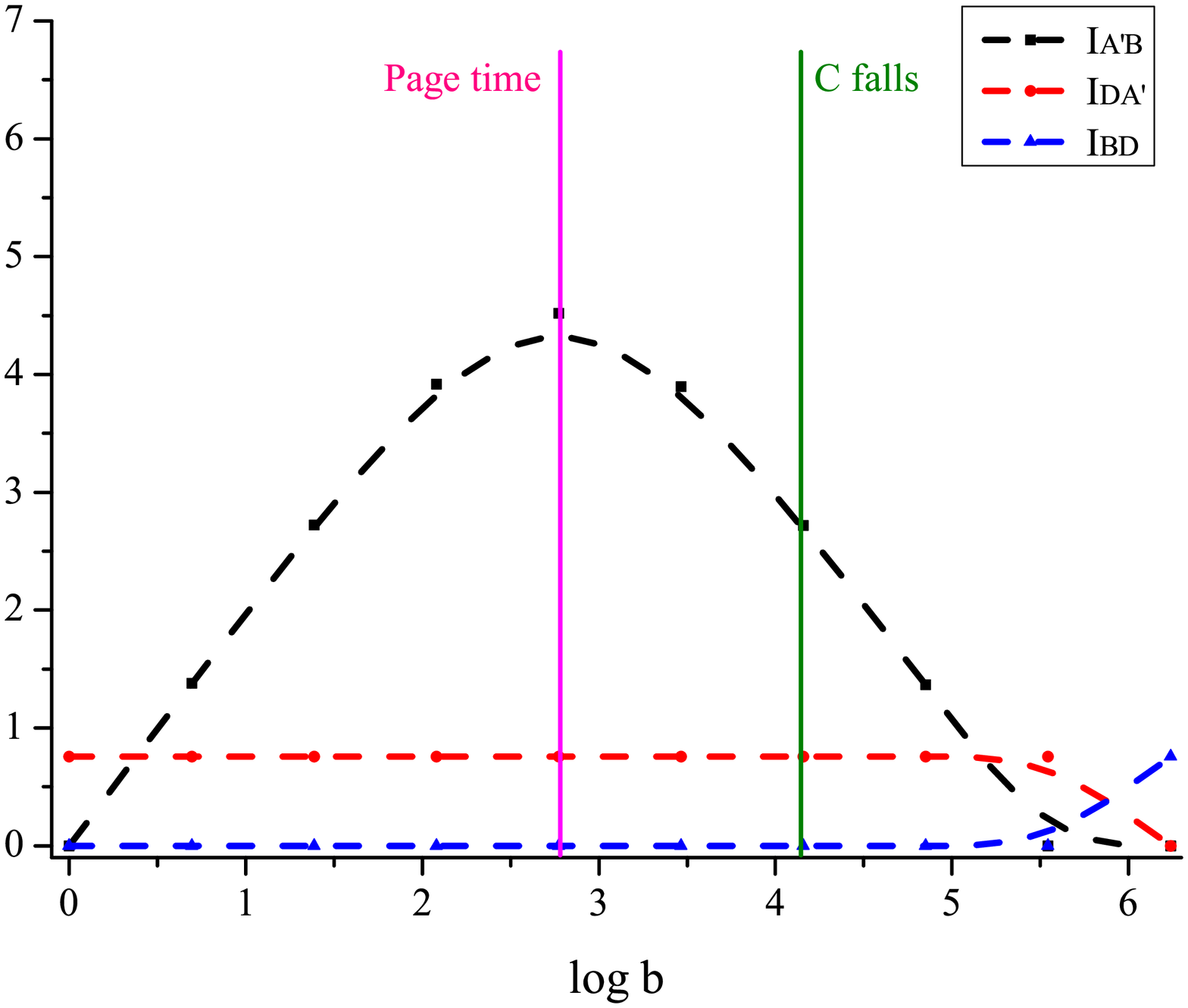}
\includegraphics[scale=0.4]{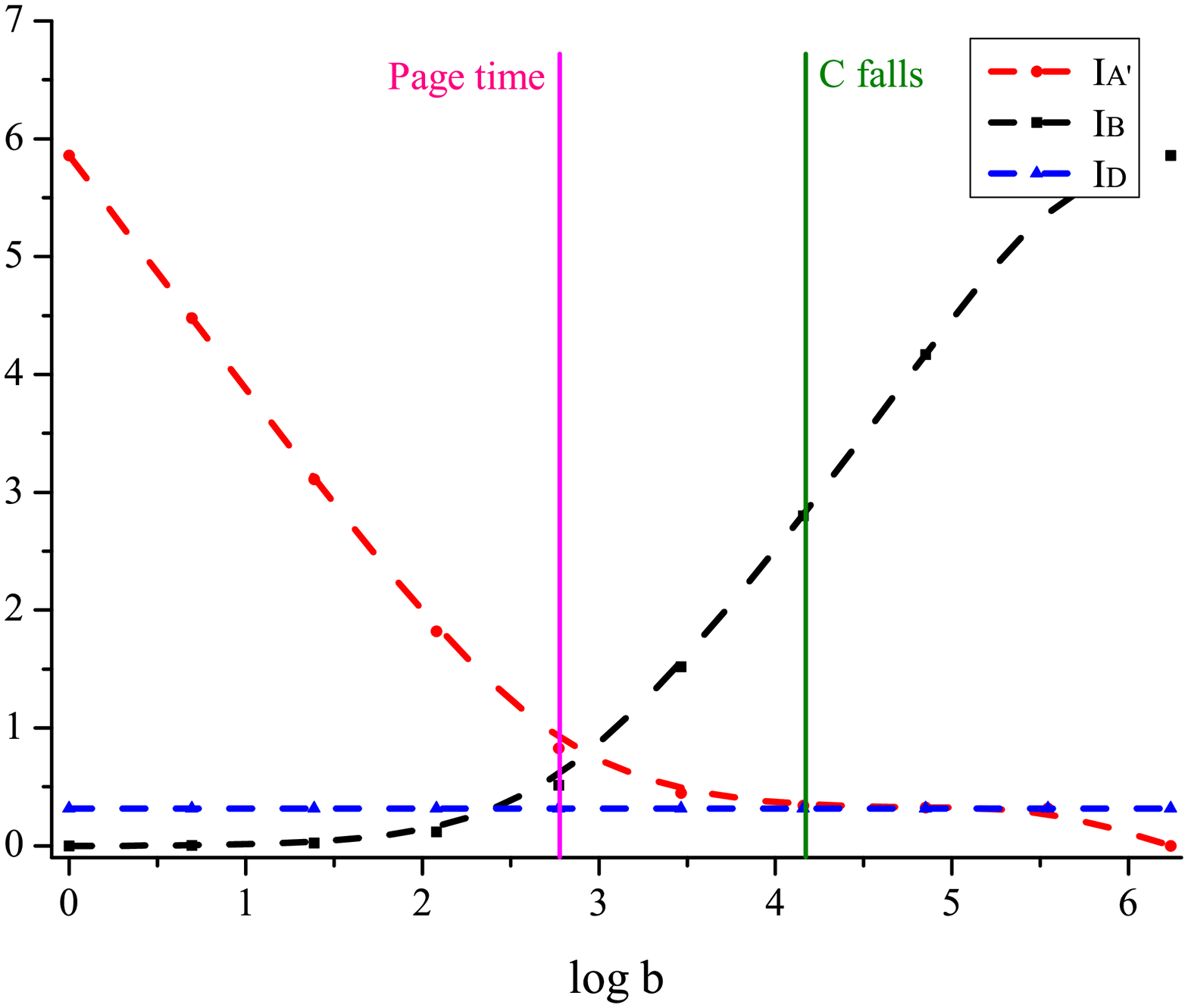}
\caption{\label{fig:data120}Same condition as in Fig.~\ref{fig:data12}, but without a randomization process.}
\end{center}
\end{figure}

Keeping this in mind, we report the result. The general tendency is consistent with the previous result without $C$ and $D$. The critical observation is that soon after $C$ falls into the black hole, $I_{BD}$ (blue dashed curve of the upper figure of Fig.~\ref{fig:data12}) increases to its maximum value, which is no longer a trivial result. This is due to the randomization between $A$ and $C$. If we turn off the swap gate, then the result is Fig.~\ref{fig:data120}, which clearly shows that no emission via $I_{BD}$ occurs until the last stage. Therefore, with no randomization process, no mirror-like emission of information can occur, as was expected by Hayden and Preskill.

\subsection{Importance of the Sufficient Mixing}

\begin{figure}
\begin{center}
\includegraphics[scale=0.4]{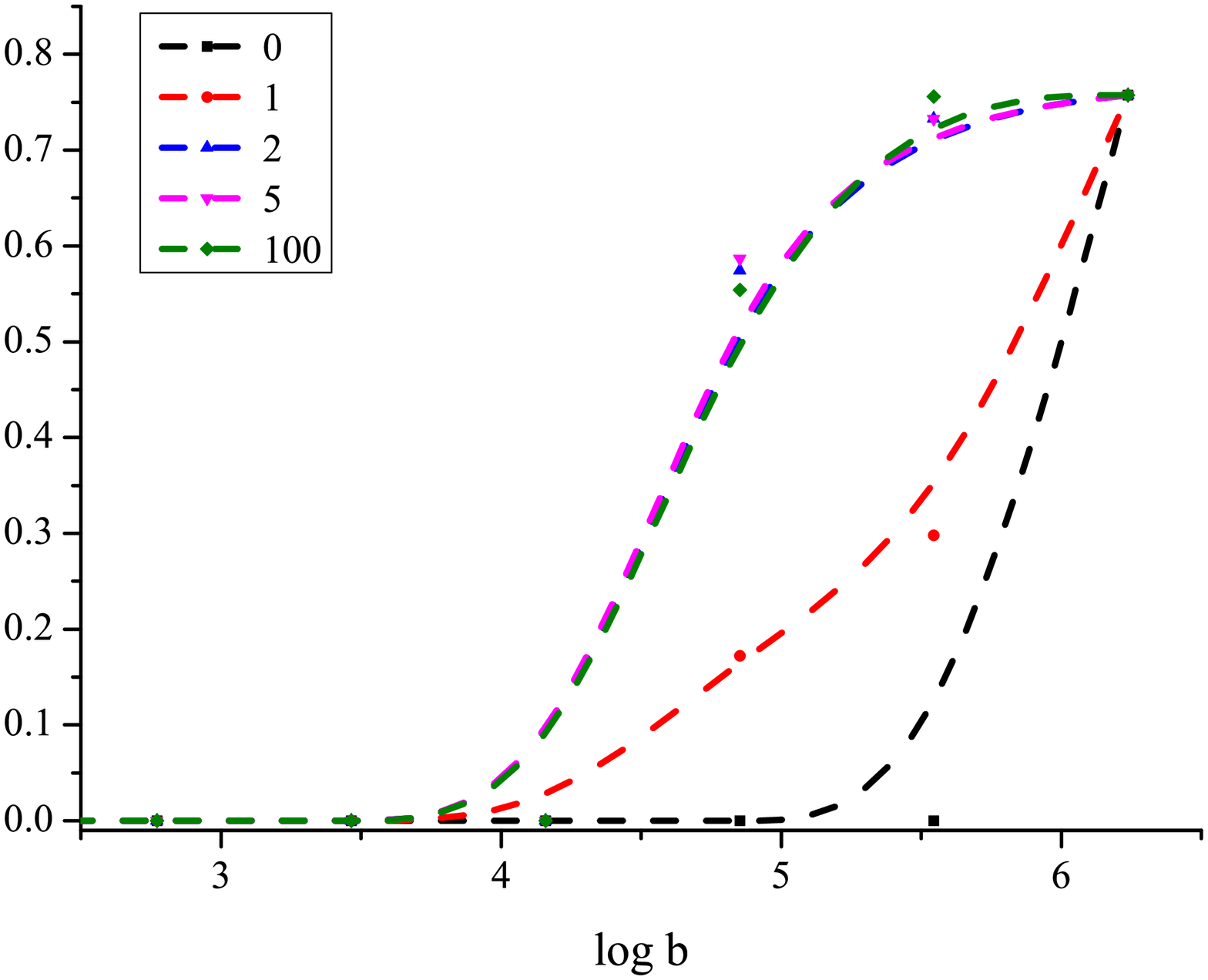}
\caption{\label{fig:mixing}$I_{BD}$ for varying number of mixing operation, say $0, 1, 2, 5,$ and $100$, respectively.}
\end{center}
\end{figure}

By varying the number of swap gate operations, we can see the tendency of the rapid emission via mutual information. Fig.~\ref{fig:mixing} clearly shows a dependence on the number of mixing operations. If it is zero, then no emission via mutual information occurs. If it is only one and, hence, is insufficient, then the emission carries a little bit of information via mutual information, but not fast enough. If the operation is repeated more, then very quickly, the curve becomes saturated. Thus, we can conclude that rapid mixing is significant for the Hayden-Preskill protocol and that the swap gate operation is very effective in ensuring the sufficient mixing of entanglements.

Due to limitations of the computation, we could not extend the number of particles more. However, as we increase the number of particles more and develop a better calculation technique, we may see a clearer relation between mixing and information flow. We leave this topic for a future investigation.

\section{\label{sec:con}Conclusion}

In this paper, we constructed the Hayden-Preskill protocol by using spin-1/2 particles and checked numerically whether it functionally works or not. To do this, we introduced four parts corresponding to the black hole $A$, the radiation $B$, the infalling information $C$, and its entangled counterpart $D$. After the Page time, if $C$ is inserted and if $A$ and $C$ are mixed sufficiently, then the information about $C$ rapidly escapes to $B$ in the form of the mutual information $I_{BD}$. Therefore, we numerically demonstrated that the Hayden-Preskill protocol can be realized using our toy system.  

One of the remarkable observations is the importance of rapid mixing between $A$ and $C$. If no mixing occurs, then emission is not rapid. After $A$ and $C$ are rapidly mixed, one needs to take out a particle from $A \cup C$. In real black holes, rapid mixing is not a strange assumption. However, the question is, \textit{where is the place of the rapid mixing}. If the place is inside the horizon (hence, near the central singularity), then as many authors are claiming, particles cannot be directly taken from $A \cup C$ to $B$ after the mixing of $A$ and $C$; if the particle emission is generated by a separable particle-antiparticle pair, then the total number of degrees of freedom should purely increase, and no emission of information will take place after the Page time as we observed in \cite{Hwang:2017yxp}. On the other hand, if one follows the membrane paradigm, the black hole complementarity principle, or the firewall picture \cite{Almheiri:2012rt}, then rapid mixing happens at the horizon; hence, taking one particle out from $A \cup C$ would not be surprising.

Note that measurement of the mutual information between the radiation $B$ and the counterpart $D$ is experimentally possible, in principle, even though we do not fall into the black hole. Although this is still hypothetical and depends on various assumptions, we can carefully argue that an experimental test of the Hayden-Preskill protocol being doable only outside the black hole might be a way to test whether a firewall or a membrane exists near the horizon or not. In the astrophysical context, the only possible candidate is a primordial black hole, although the real investigation may be too difficult. However, Hawking-like radiation and information flow are well defined in several examples of analog black holes \cite{Chen:2017lum}. The application of the Hayden-Preskill protocol to analog black hole models can be an interesting future research topic.

If we can prove experimentally that no firewall exists near the horizon, then either we need to accept the loss of information or an alternative explanation \cite{Sasaki:2014spa}. We hope to find a way for this to be done experimentally, at least, in principle.

\begin{acknowledgments}
DY was supported by the National Research Foundation of Korea (Grant No.: 2018R1D1A1B07049126). This work was supported by the Daegu Gyeongbuk Institute of Science and Technology (DGIST) Undergraduate Group Research Project (UGRP) grant. 
\end{acknowledgments}

\end{document}